\documentclass[twocolumn,floatfix,tighten]{aastex62}

\received{}
\revised{}
\accepted{}
\submitjournal{ApJ}

\usepackage{graphicx}	
\usepackage{amsmath}	
\usepackage{amssymb}	
\usepackage{bm}		
\usepackage{graphbox}
\usepackage{booktabs}
\usepackage{listings}
\usepackage{color}

\definecolor{dkgreen}{rgb}{0,0.6,0}
\definecolor{gray}{rgb}{0.5,0.5,0.5}
\definecolor{mauve}{rgb}{0.58,0,0.82}

\let\oldAA\AA
\renewcommand{\AA}{\text{\normalfont\oldAA}}

\usepackage[T1]{fontenc}
\usepackage{ae,aecompl}

\usepackage{newtxtext,newtxmath}

\usepackage{xcolor}
\definecolor{alizarin}{rgb}{0.82, 0.1, 0.26}

\shortauthors{Schaefer et al.}
\shorttitle{N/O - O/H and metallicity gradients}
\begin{document}

\title{SDSS-IV MaNGA: Variations in the N/O - O/H relation bias metallicity gradient measurements}

\author{Adam L. Schaefer}
\altaffiliation{aschaefer@astro.wisc.edu}
\affiliation{Department of Astronomy, University of Wisconsin-Madison, 475N. Charter St., Madison, WI 53703, USA}

\author{Christy Tremonti}
\affiliation{Department of Astronomy, University of Wisconsin-Madison, 475N. Charter St., Madison, WI 53703, USA}

\author{Francesco Belfiore}
\affiliation{European Southern Observatory, Karl-Schwarzschild-Str. 2, Garching bei M{\"u}nchen, 85748, Germany}

\author{Zachary Pace}
\affiliation{Department of Astronomy, University of Wisconsin-Madison, 475N. Charter St., Madison, WI 53703, USA}

\author{Matthew A. Bershady}
\affiliation{Department of Astronomy, University of Wisconsin-Madison, 475N. Charter St., Madison, WI 53703, USA}
\affiliation{South African Astronomical Observatory, P.O. Box 9, Observatory 7935, Cape Town, South Africa}

\author{Brett H. Andrews}
\affiliation{University of Pittsburgh, PITT PACC, Department of Physics and Astronomy, Pittsburgh, PA 15260, USA}

\author{Niv Drory}
\affiliation{McDonald Observatory, The University of Texas at Austin, 1 University Station, Austin, TX 78712, USA}

\begin{abstract}

In this paper we use strong line calibrations of the N/O and O/H in MaNGA spaxel data to explore the systematics introduced by variations in N/O on various strong-line metallicity diagnostics. We find radial variations in N/O at fixed O/H which correlate with total galaxy stellar-mass; and which can induce $\sim 40 \%$ systematic uncertainties in oxygen abundance gradients when nitrogen-dependent abundance calibrations are used.
Empirically, we find that these differences are associated with variation in the local star formation efficiency, as predicted by recent chemical evolution models for galaxies, but we cannot rule out other processes such as radial migration and the accretion of passive dwarf galaxies also playing a role.

\end{abstract}

\section{Introduction}
The accurate determination of galactic elemental abundances provides a powerful tool for exploring the assembly and evolution of galaxies into the structures that we see today. Gas-phase oxygen is produced in massive stars and redistributed into the interstellar medium (ISM) by Type II supernovae on relatively short ($\sim 10 \, \mathrm{Myr}$) timescales following an episode of star formation. The concentration of oxygen in the ISM of a galaxy, and spatial gradients in this quantity, are therefore sensitive to a number of physical processes, including gas inflows, gas outflows and the rate of star formation \citep[e.g.][]{Lilly2013, Belfiore2019b}. The production of nitrogen in galaxies is a more complicated process. Unlike oxygen, nitrogen is both a primary and a secondary element, meaning that at high metallicity the total quantity produced following a burst of star formation is dependent on the initial abundance of carbon and oxygen \citep{Matteucci1986,VilaCostas1993,Vincenzo2016}. A small amount of primary nitrogen is produced directly from the fusion of lighter elements, but much more is produced during the post-main-sequence evolution of low and intermediate mass stars through the CNO cycle. Secondary production of nitrogen dominates in stars with O/H above $\sim 0.25$ times the solar value \citep{Nicholls2017}.

Since the synthesis of nitrogen and oxygen occur in stars of different masses and with different lifetimes, the enrichment of the interstellar medium with these elements occurs on different timescales as well, with nitrogen continuing to be released into the ISM for over a Gyr after an initial burst of star formation \citep[][but attributed to Vincenzo \emph{in prep.} therein]{Maiolino2019}. This means that with varying conditions of star formation in a galaxy, the ratio of N/O at a given O/H can vary significantly. For example, if the rate of star formation is very high, oxygen is produced by Type II supernovae very rapidly, but the production of nitrogen will lag behind. This has been observed in some high-redshift starburst galaxies, where N/O at a fixed O/H can be up to three times lower than in galaxies with more modest star formation rates \citep[e.g.][]{Pettini2002}. Conversely, galaxies with a low star formation efficiency will produce relatively small amounts of oxygen on short timescales while the creation of nitrogen by previous generations of stars continues, driving N/O to higher values \citep{Molla2006,Vincenzo2016}. Moreover, the relative yields of nitrogen and oxygen are affected by the stellar initial mass function (IMF) with which stars are formed, with bottom-heavy IMFs producing a larger proportion of nitrogen. Further variation in the N/O ratio within galaxies is possible through changes in the accretion timescale and the chemical abundances in the accreted material. 

Previous work has shown that in galaxies in the low-redshift universe, there are systematic radial variations in several of the quantities that can determine the N/O ratio at fixed O/H. \cite{Leroy2008} showed that the star formation efficiency (defined as the ratio between the star formation rate and the total molecular plus atomic gas mass) in spiral galaxies can vary by up to a factor of $10$ between their centres and their outskirts, with the centres generally hosting more efficient star formation than at large radii. Using optical integral fields spectroscopic observations, \cite{Parikh2018} found that early-type galaxies showed evidence for radial variation in the IMF, with a higher proportion of low-mass stars expected to form in the centres of galaxies. With both of these processes potentially at play, the possibility of systematic variations in the N/O versus O/H relation within and between galaxies is real. This possibility was corroborated by \cite{Belfiore2017}, who found that at fixed oxygen abundance the N/O ratio in more massive galaxies is higher than in less massive galaxies. Noting that the regions in massive galaxies that have the same oxygen abundance as less massive galaxies occur at very different galactocentric radii, the authors of this study speculated that radial variations in the star formation efficiency (SFE) could explain the observed trends.

The secondary nature of nitrogen production is a feature that has been exploited for numerous strong-line oxygen abundance diagnostics. Since N/O is proportional to the O/H at high metallicity, it is possible to estimate the total oxygen abundance using emission line diagnostics that trace the N/O ratio. This assumption of a fixed relation is made explicitly for some metallicity indicators such as the N2O2 calibrator of \cite{Kewley2002} or the N2S2H$\alpha$ calibrator of \cite{Dopita2016} that were derived from photoionisation modelling, or implicitly for some empirically calibrated indicators such as O3N2 from \cite{Pettini2004}. As a consequence, variation of the relationship between N/O and O/H will have an effect on the oxygen abundance inferred from a given set of emission line ratios. Given the observations of \cite{Belfiore2017}, the derivation of metallicity gradients with some strong line methods may be susceptible to biases induced by differences between the conditions in galaxy centres and their outer parts.

In this letter we further investigate the radial variations of the N/O versus O/H relation, and study the impact of this variation on the oxygen abundance gradients inferred from strong line calibrations that include an emission line from nitrogen. We will assume a flat $\Lambda$CDM cosmology with $H_{0} = 70 \, \mathrm{km \, s^{-1} \, Mpc^{-1} }$, $\Omega_{m}=0.3$ and $\Omega_{\Lambda} = 0.7$, and a \cite{Chabrier2003} IMF unless otherwise stated.

\section{Data and Methods}
To perform our investigation of the relative abundances of nitrogen and oxygen, we make use of the rich, spatially-resolved spectroscopic dataset provided by the Mapping Galaxies at Apache Point Observatory  galaxy survey \citep[MaNGA;][]{Bundy2015}, which is a component of SDSS-IV \citep{Blanton2017,Abolfathi2018}. This survey, performed on the $2.5 \, \mathrm{m}$ SDSS telescope at Apache Point Observatory \citep{Gunn2006}, uses an array of $17$ hexabundles composed of between $19$ and $127$ optical fibres \citep{Drory2015}. These integral field units observe the MaNGA primary sample out to $1.5 \, R_{e}$ and the higher redshift secondary sample out to $2.5 R_{e}$ on average \citep{Wake2017}. Light from these units is fed into the BOSS spectrograph \citep{Smee2013} where it is dispersed to a resolution of $R \sim 2000$ over wavelengths in the range $3600 - 10300 \, \mathrm{\AA}$ \citep{Law2015,Yan2016}.

For this work, we draw our sample from $8$th MaNGA Product Launch (MPL-8), which includes data for $6507$ unique galaxies.
Measurements of the emission line fluxes in each spaxel are obtained from the MaNGA Data Analysis Pipeline \citep{Westfall2019,Belfiore2019}, which fits single-component Gaussians to the emission lines in the reduced data cubes \citep{Law2016}. Following \cite{Schaefer2019}, we select only galaxies that are close to face-on by discarding galaxies with minor to major axis ratio, $b/a<0.6$. We further reject galaxies that have fewer than $60\%$ of their observed spaxels in the star-forming region of the \cite{Baldwin1981} diagram. Furthermore, galaxies with their $r$-band effective radius smaller than $4\arcsec$ are rejected to minimise the effect of the observational point spread function on the estimation of gradients in our sample. For a full discussion and justification of these choices, see \cite{Schaefer2019} and \cite{Belfiore2017}. Our final sample comprises $1008$ star-forming galaxies.

\subsection{Dust extinction corrections}
The emission lines that will be used for our analysis span a wide range of wavelengths, and as such their observed intensities are affected by dust attenuation along our line of sight. We estimate the reddening by comparing the observed ratio of H$\alpha$ to H$\beta$ to the theoretical case B value at $10^{4} \, \mathrm{K}$ of $2.86$. Extinction for each emission line is then corrected for by assuming a \cite{Cardelli1989} dust law with the ratio of total to selective extinction $R_{V} = 3.1$. In what follows, all measurements of line fluxes have been dust-corrected in this manner.

\subsection{O/H}
We wish to understand the relationship between the relative abundances of nitrogen and oxygen, and the impact of these on our oxygen abundance determinations. To this end, we make two measurements of the oxygen abundance: one that does not incorporate the flux of a nitrogen line that will induce an artificial correlation with the nitrogen abundance, and one that does. 

\subsubsection{R23 oxygen abundance}
For our nitrogen-free indicator we use the R23 calibration of \cite{Maiolino2008}. This uses a combination of the [OII]$\lambda3726,3729$, [OIII]$\lambda 4959,5007$ and H$\beta$ emission lines,
\begin{equation}
\mathrm{R23} = \frac{\mathrm{[O III]\lambda 5007, 4959 + [O II]\lambda 3726,3729} }{\mathrm{H\beta}}.
\end{equation}

At low oxygen abundance, this indicator was calibrated against a collection of local low-metallicity HII regions with auroral [OIII]$\lambda 4363$ detections. At high metallicity, it was calibrated against a sample of $22,482$ galaxies from the SDSS spectroscopic sample, with metallicities determined from photoionization modeling.
With their data, \cite{Maiolino2008} found that the R23 ratio is well described by a 4th degree polynomial function of metallicity,

\begin{equation}
\log(\mathrm{R23}) = 0.7462 - 0.7149 x - 0.9401 x^{2} - 0.6154 x^{3} - 0.2524 x^{4},
\end{equation}
where $x$ is O/H relative to the assumed solar value of ($x=12+\log(\mathrm{O/H}) - 8.69$), assuming the solar value derived by \cite{Asplund2005}. To compute the metallicity, we measure R23 and then solve the polynomial. While this polynomial is double-valued, the vast majority of spaxels in MaNGA are on the upper branch. Errors on R23 were calculated by a Monte Carlo method where the observed fluxes were deviated from their measured value by a Gaussian with $\sigma$ equal to the observed uncertainties. This process was repeated $500$ times and the standard deviation of the resulting distribution is taken to be the measurement error on $12+\log(\mathrm{O/H})$.

\subsubsection{[OIII]/[NII] oxygen abundance}
In addition to the nitrogen-free R23 calibrator, we also estimate the oxygen abundance by employing the commonly used combination of [NII]$\lambda 6584$ and [OIII]$\lambda 5007$ lines\footnote{This indicator is very similar to the \cite{Pettini2004} O3N2 calibration, though the metal lines are not taken in ratio with the wavelength-adjacent Balmer lines, meaning that a dust attenuation correction is required.}. The calibration of the ratio of these two lines to derive oxygen abundances was also performed by \cite{Maiolino2008} using the same data and model grids as for the R23 indicator. Using this method we ensure that both of our oxygen abundance measurements have the same absolute abundance scalings and can therefore be easily compared. \cite{Maiolino2008} found that the ratio $\mathrm{R}=\mathrm{[OIII]/[NII]}$ is adequately fitted by a third degree polynomial 
\begin{equation} 
\log(\mathrm{R}) =  0.4520 - 2.6096 x - 0.7170x^{2} + 0.1347 x^{3}, 
\end{equation}
where again $x=12+\log(\mathrm{O/H}) - 8.69$. As with the R23-based O/H calibration, we derive uncertainties on the individual measurements using a Monte Carlo method based on $500$ realisations of the data with Gaussian noise added based on the line flux uncertainties calculated by the {\tt DAP}. The inclusion of the [NII]$\lambda6584$ line in this oxygen abundance determination means that this indicator will be sensitive to variation in N/O at fixed O/H.

\subsection{N/O}
To estimate the abundance of nitrogen relative to oxygen, we make use of the method presented by \cite{Thurston1996}. This procedure utilises the ratio of [NII]$\lambda 6548,6584$ to [OII]$\lambda 3726,3729$. The N/O ratio is derived from the emission line fluxes and the temperature of the NII ions, $t_{II}$, using a formula obtained from a theoretical five-level atom calculation,

\begin{equation} \label{NO_eq}
\log(\mathrm{N/O}) = \log\left(\frac{\mathrm{[NII]\lambda 6548,6584}}{\mathrm{[OII]\lambda 3726,3729}} \right) + 0.307 - 0.02 \log( t_{\mathrm{II}} )- \frac{0.726}{t_{\mathrm{II}}}. 
\end{equation}
For our purposes, we derive the temperature using the R23 ratio,
\begin{equation}
t_{\mathrm{II}} =  6065 + 1600 \log (\mathrm{R23}) +1878 (\log \mathrm{R23} )^{2} + 2803 (\log \mathrm{R23})^{3},
\end{equation}
which was derived from a set of photionisation models by \cite{Thurston1996}. The estimation of the N/O ratio suffers from smaller systematics than the estimation of O/H since the ionisation potential of N+ and O+ are very similar ($14.53 \, \mathrm{eV}$ and $13.61 \, \mathrm{eV}$ respectively).

Note that at fixed R23, Equation \ref{NO_eq} implies a one-to-one correspondence between $\log(\mathrm{N/O})$ and $\log(\mathrm{[NII]/[OII]})$. That is to say, a factor of two increase in the relative abundance of nitrogen would result in a factor of two increase in the relative strength of the [NII] lines.

The estimation of N/O relies on a temperature-dependent correction based on R23, which is also used to derive the nitrogen-free metallicity. This sets the absolute scaling of N/O at fixed O/H, but also induces a correlation between the two measurements that is largest at highest oxygen abundance. Fortunately the form of the correction is such that $\log(\mathrm{[NII]/[OII]})$ is added to the terms that include R23. This ensures that at fixed R23 (fixed oxygen abundance) variation in N/O is driven almost independently by changes in the [NII]/[OII] flux ratio, and a valid comparison of N/O at fixed O/H in different galaxies can be made.

\section{Results}
\begin{figure}
\includegraphics{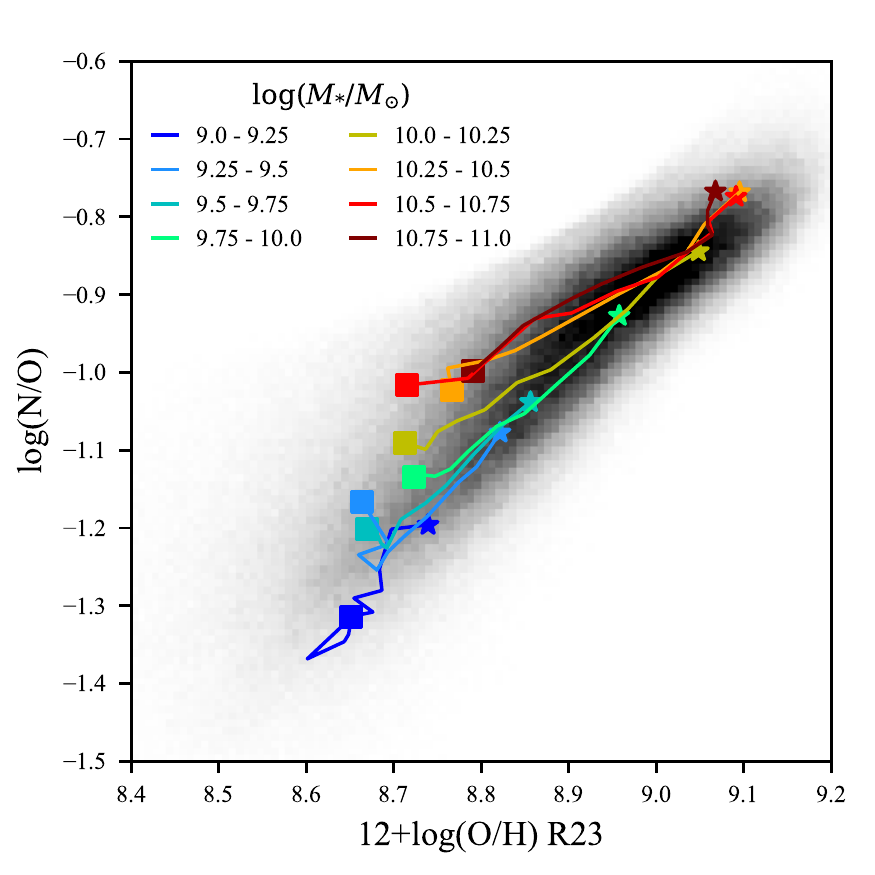}
\caption{The distribution of N/O and O/H in our sample. The greyscale shows the density of measurements from individual spaxels. The coloured lines are the median profiles of $\log(\mathrm{N/O})$and $12+\log(\mathrm{O/H})$ evaluated in $0.2 \, R_{e}$-wide radial bins for galaxies in the stellar mass range indicated by the colour of the line. The central parts of the galaxies are indicated by the star symbols and the square symbols are at $2 \, R_{e}$. While the outer parts of massive galaxies have the same oxygen abundance as the inner parts of low-mass galaxies, the more massive galaxies exhibit a higher abundance of nitrogen relative to oxygen.}\label{OH_NO_RPs}
\end{figure}

With the measurements in place we are now in a position to explore the differences between our strong-line O/H indicators and the relationship between these differences and the relative abundance of nitrogen. In Figure \ref{OH_NO_RPs} we show the overall distribution for N/O and O/H in greyscale. Overplotted in colour are the radial trends in N/O and O/H separtated by total galaxy stellar mass. To compute these median trends, we calculated the mean $\log(\mathrm{N/O})$ as a function of radius for individual galaxies, then calculated the median for all galaxies in $0.2 \, R_{e}$-wide radial bins. Similar to \cite{Belfiore2017}, the gradient of these curves in this parameter space differs for galaxies of different stellar masses. The more massive systems appear to have higher N/O than less massive systems at fixed O/H, and the inner parts of low-mass galaxies do not have the same abundance patterns as the outer parts of high-mass galaxies. This difference manifests in as much as a $0.2 \, \mathrm{dex}$ offset in N/O at fixed O/H between the most and least massive galaxies in our sample.

\begin{figure*}
\begin{center}
\includegraphics{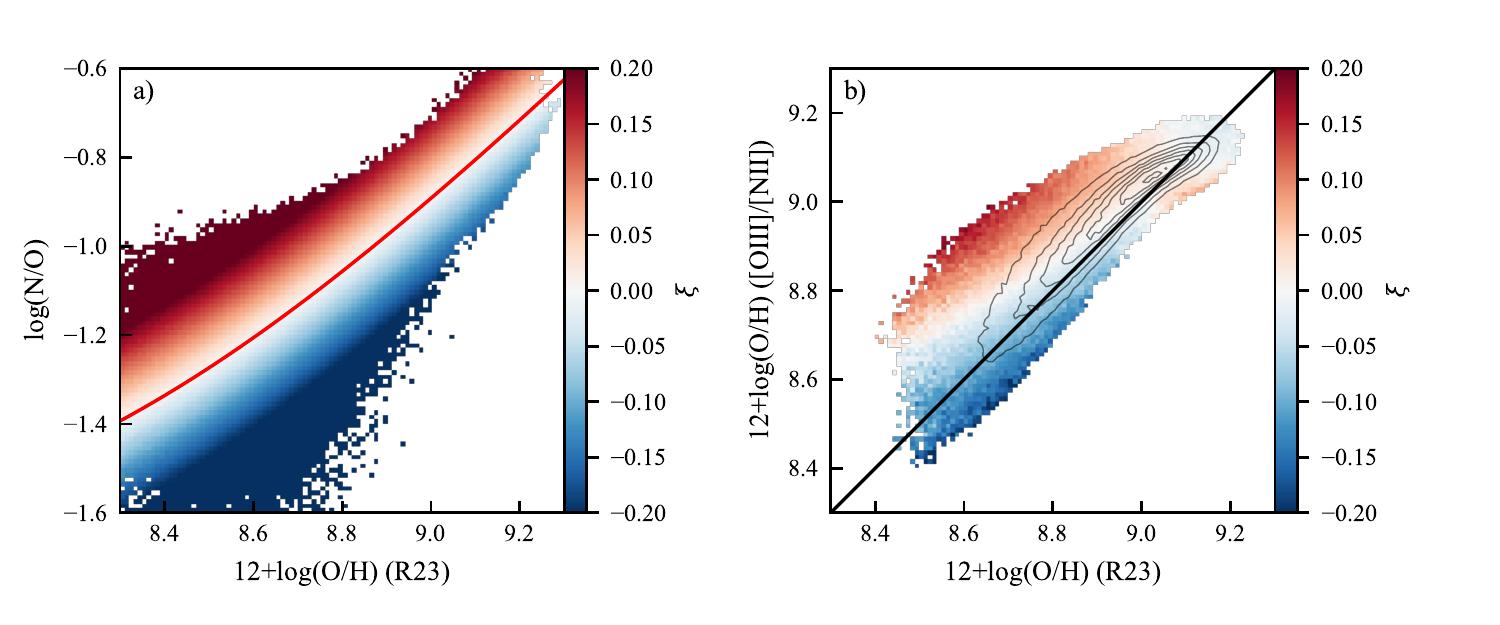}
\caption{ A comparison of two oxygen abundance indicators, and the correlation between the differences with the nitrogen excess factor, $\xi$. In panel a) we illustrate the definition of $\xi$ as the excess or deficiency of $\log(\mathrm{N/O})$ above or below the expected value, marked in red. In panel b) we plot the [OIII]/[NII]-derived metallicity against the R23-based calibration. The overall distribution of points is traced by the grey contours, while the colour corresponds to the mean $\xi$ for data points calculated in each pixel in the image. We include pixels only where $20$ or more data points can be used to calculate the mean. The black line is the one-to-one relation. Where the [OIII]/[NII] oxygen abundance is higher than the R23 estimation, the observed $\xi$ is also higher.}\label{NO_excess}
\end{center}
\end{figure*}

The systematic difference in the N/O versus O/H scaling relation between galaxies of different stellar mass will have a systematic effect on some oxygen abundance calibrations. In Figure \ref{NO_excess} we explore how variation in the nitrogen abundance at fixed oxygen abundance impacts the calibration based on [OIII]/[NII]. To do so, we first define the nitrogen excess factor, $\xi$. Following \cite{Nicholls2017} we fit the N/O versus O/H data with a function of the form $\log(\mathrm{N/O}) = \log(10^{a} + 10^{\log(\mathrm{O/H}) + b})$, where $a$ describes the primary production of nitrogen, and $b$ describes the oxygen-sensitive secondary production at higher metallicity. As we have very little low-metallicity data to constrain the primary N/O ratio, we take the \cite{Nicholls2017} value of $a = -1.732$, and find $b=2.041$ with our fit. This line is marked in red in panel a) of Figure \ref{NO_excess}. We then define $\xi$ as the $\log(\mathrm{N/O})$ residual from this line. 

In panel b) of Figure \ref{NO_excess} we compare the [OIII]/[NII] metallicity to the R23-based indicator. While these two measures have been calibrated from the same data and models, they do not match perfectly. The density of data points is shown with the grey contours and the colour in each pixel of the image represents the mean $\xi$ for data points in that cell. This map shows that differences in the O/H measurements are highly correlated with the excess of nitrogen above or below the expected value.

\begin{figure*}
\includegraphics{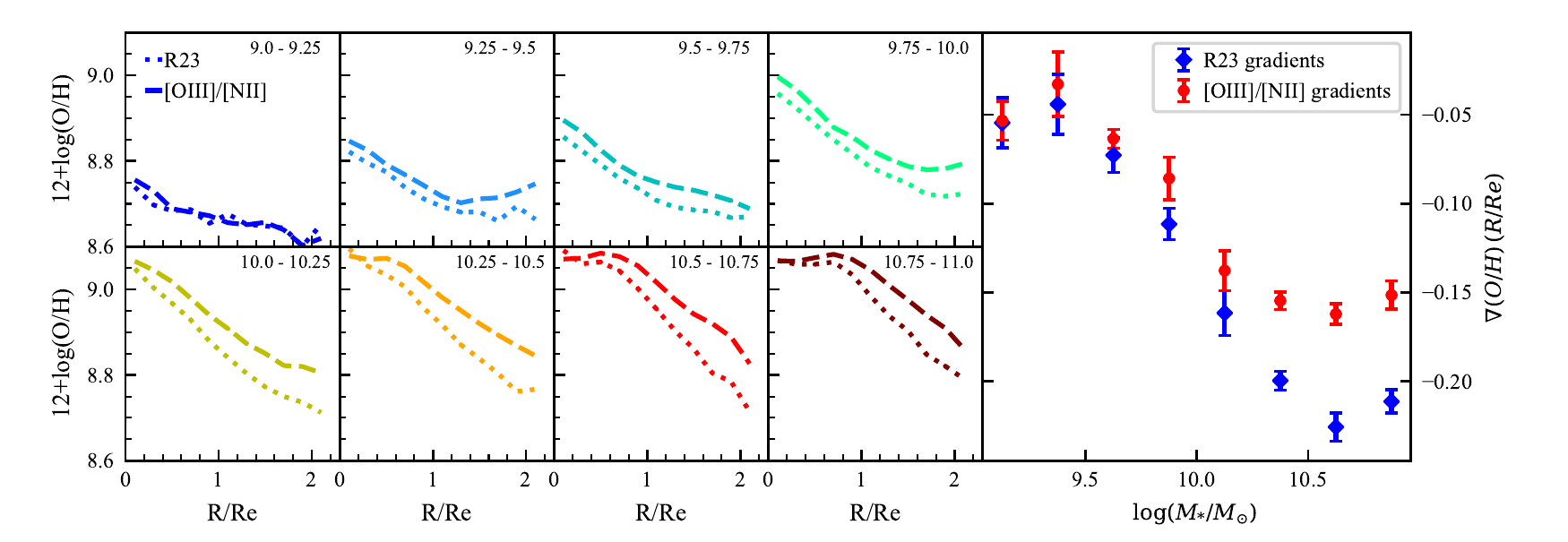}
\caption{Discrepancies in the inferred radial profiles of oxygen for the two different abundance indicators. The small panels show the radial profiles of $12+\log(\mathrm{O/H})$ as derived by R23 (dotted) and [OIII]/[NII] (dashed). Profiles are shown for galaxies within the range of stellar masses indicated at the top right of each panel. While [OIII]/[NII] gives slightly higher O/H, the gradients are also flatter. The gradients are shown as a function of stellar mass in the large panel on the right. The difference between the gradients yielded by the two indicators grows with stellar mass, culminating in a disagreement of approximately $40$ per cent  above $\log(M_{*}/M_{\odot}) > 10.5$.}\label{Gradient_differences}
\end{figure*}

The systematic difference in N/O at fixed O/H that exists between galaxies of different stellar masses as shown in Figure \ref{OH_NO_RPs} will influence the radial profiles of $12 + \log(\mathrm{O/H})$ inferred from different strong-line abundance indicators. We explore this by measuring the oxygen abundance gradients with both R23 and [OIII]/[NII]. To do so, we construct the median radial profile using the methodology outlined in \cite{Schaefer2019}. We then measure the gradient by fitting a linear least squares fit to the median profile in the radial range $0.5<R/R_{e} < 2$. These profiles are shown in Figure \ref{Gradient_differences}, along with the measured gradients. The O/H gradient is seen to steepen with increasing stellar mass, in agreement with \cite{Belfiore2017}. We find that the abundance measurements differ between the two calibrators, with a systematic difference in the gradients that increases with the integrated stellar mass. The measured slopes are almost identical at smaller radii, but the discrepancy increases to $0.06 \, \mathrm{dex} \, R_{e}^{-1}$ above $M_{*} = 10^{10.5} \, \mathrm{M_{\odot}}$. Given that the magnitude of the [OIII]/[NII] abundance gradients in this mass range are $\sim -0.15$, this represents a systematic difference of $\sim 40$ per cent.

\section{Discussion and Conclusion}
Strong-line metallicity calibrators are used ubiquitously throughout the study of observational galaxy evolution. Many of these calibrations operate by assuming a fixed relationship between N/O and O/H. In these schemes, the strength of the emission lines of a primary element such as oxygen or sulphur are compared to the strength of the nitrogen line, a secondary element, and from this an oxygen abundance is derived. We have shown that the use of strong-line oxygen abundance calibrators that rely on this fixed scaling of primary to secondary elements can result in unexpected systematic effects that may have a substantial impact on the conclusions of some papers.

\subsection{The physical origin of the N/O offset}
Systematic differences in local oxygen abundances as reported by different indicators appear to have a radial dependence. That is to say, the regions of more massive galaxies that are observed to have the same O/H as lower mass galaxies are at larger radii. This is a consequence of the global mass metallicity relation and the negative metallicity gradients that tend to exist in galaxies in the mass range explored here. Models of nitrogen production in galaxies \citep[e.g.][]{Vincenzo2016} find that galaxies with lower SFE tend to have higher N/O at fixed O/H as because the birth rate of oxygen-producing massive stars is lower under these conditions. 

\begin{figure*}
\includegraphics[align=c,scale=0.8]{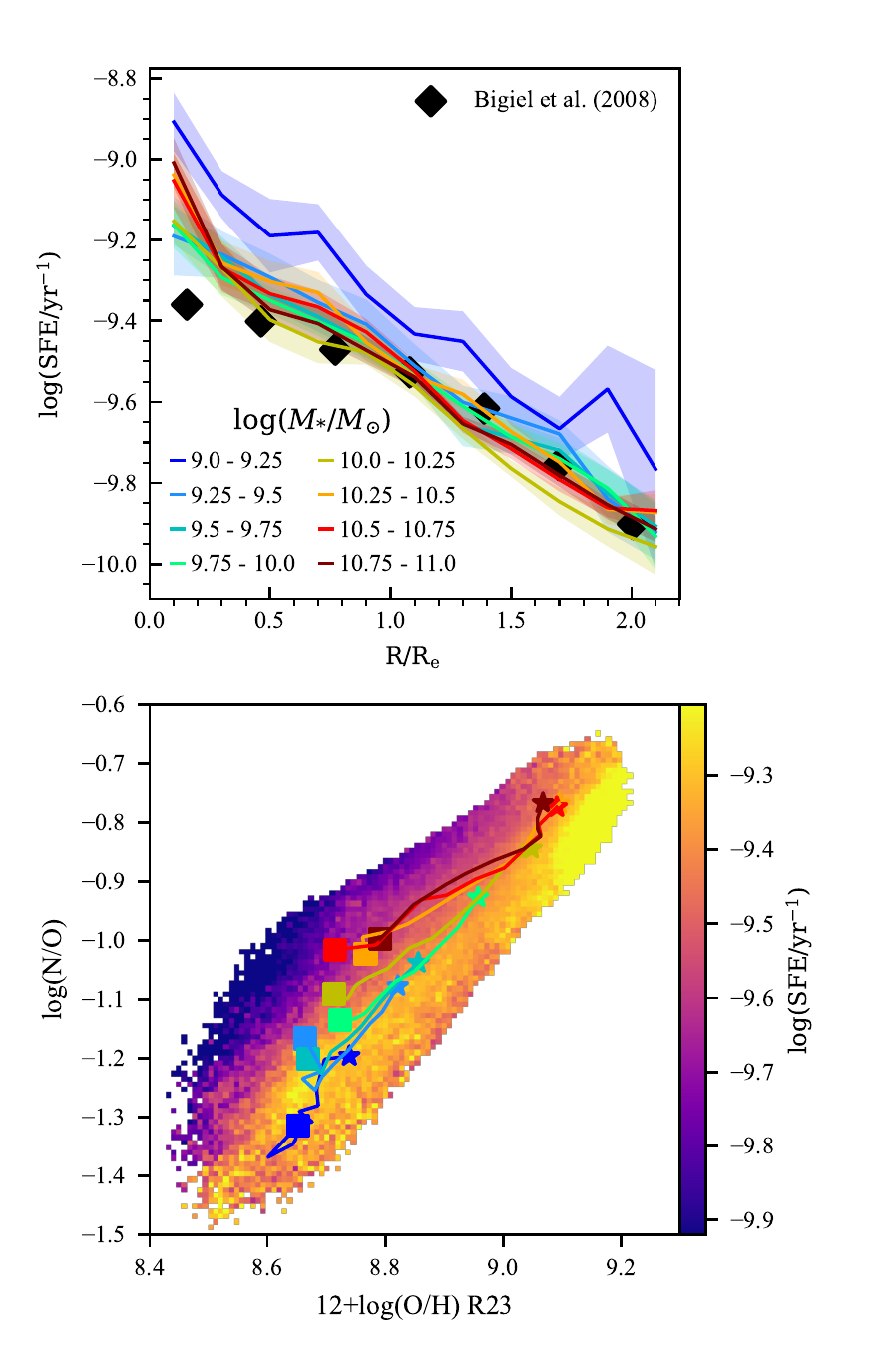}
\includegraphics[align=c,scale=0.9]{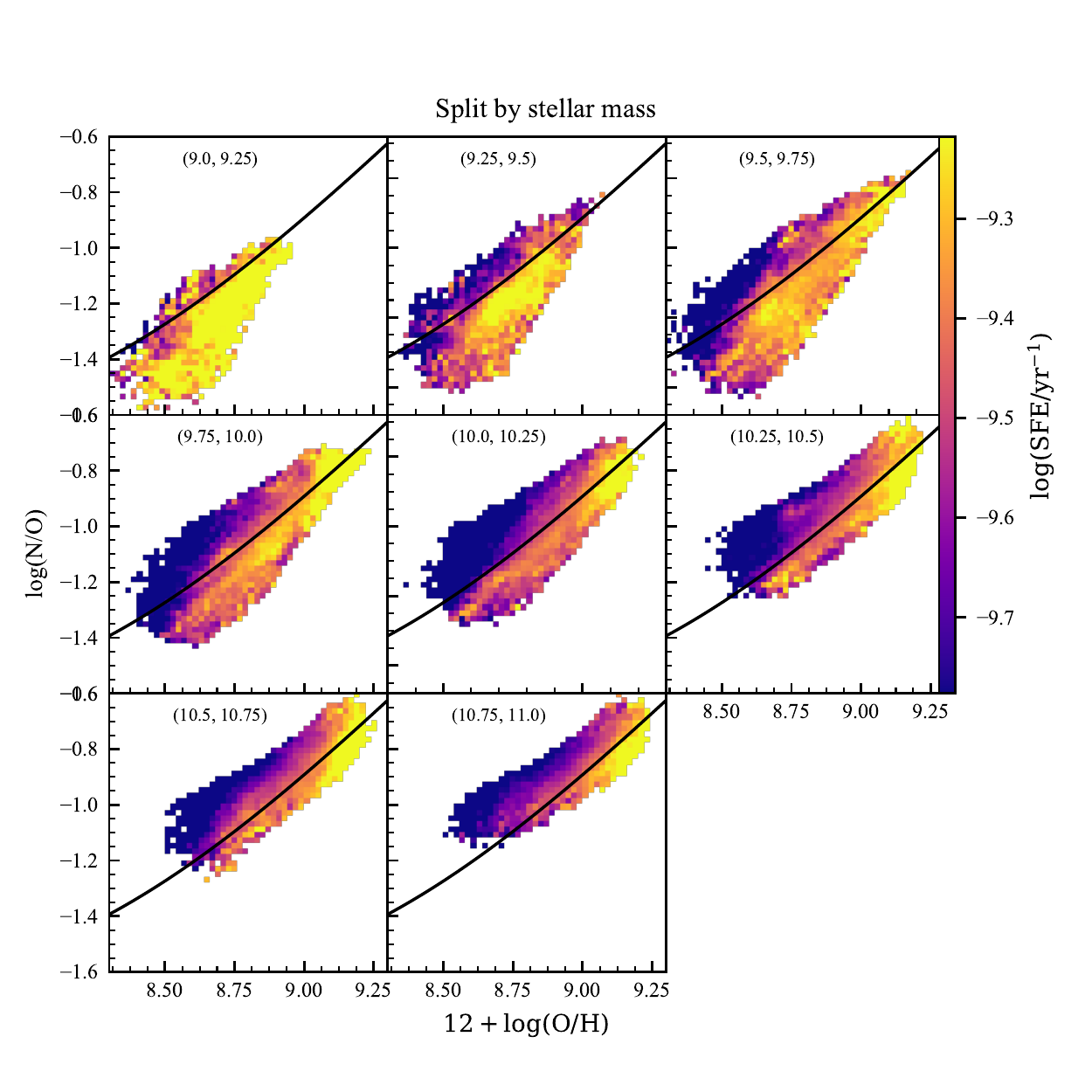}
\caption{ (\emph{Top left}) The radial distribution of star formation efficiencies estimated from the MaNGA data. These are consistent with the \cite{Bigiel2008} estimates shown by the black diamonds. (\emph{Lower left}) The distribution of $\log(\mathrm{N/O})$ and $12+\log(\mathrm{O/H})$, with colours corresponding to the mean estimated star formation efficiency for spaxels falling in each cell in the image. At fixed oxygen abundance, regions of galaxies with higher N/O exhibit systematically lower SFE. We have plotted the median radial dependence for galaxies in the stellar mass range indicated by the colour of the line. (\emph{Right}) We show distribution of N/O, O/H and the SFE, splitting the sample into bins of total galaxy stellar mass indicated at the top of each panel. The trend seen in the full sample is reflected in each mass bin, though each interval of stellar mass covers different parts of the N/O vs O/H plane. The $\xi=0$ line is marked in black. }\label{OH_NO_SFE}
\end{figure*}


Moreover, observations of local galaxies have shown that SFE is higher in the centre, and lower in the outskirts \citep{Leroy2008}. The measurement of the SFE in galaxies requires resolved observations of the star formation rate as well as the gas density. While we do not have access to resolved measurements of the gas surface density for the vast majority of galaxies in MaNGA, we can use local scaling relations to make an estimation. We do so by employing the methodology outlined in \cite{BarreraBallesteros2018} and \cite{Schaefer2019}, whereby the gas surface density is estimated from the dust attenuation and the assumption of a metallicity-dependent gas-to-dust ratio \citep{Wuyts2011}. \cite{BarreraBallesteros2019} found no strong dependence of the dust to gas ratio on the metallicity, however they note that their sample does not cover a broad range of oxygen abundances. Since our sample covers $\sim 0.75$ dex in O/H, we include the metallicity dependence in our gas density estimates, noting that it is at most a factor of $\sim 2$. The star formation rate can be estimated from the extinction-corrected H$\alpha$ luminosity \citep{Kennicutt1998}, with the SFE defined as the ratio between the two. Although both of these scaling relations are well established, the estimation of star formation efficiency from Balmer emission lines requires some testing. In the upper panel of Figure \ref{OH_NO_SFE} we show the radial profiles of the SFE determined by our method, compared to the values obtained from direct gas measurements for local galaxies by \cite{Bigiel2008}. We find reasonable quantitative and qualitative agreement with the \cite{Bigiel2008} quantities.

In the lower panel of Figure \ref{OH_NO_SFE}, we show the distribution of N/O and O/H, coloured by the average star formation efficiency. Higher N/O at fixed O/H corresponds to spaxels with lower star formation efficiency on average. The strength of the correlation between between the SFE and $\xi$ varies with the oxygen abundance. We find that the Spearman rank correlation coefficient ranges from $\rho=-0.5$ at $12+\log(\mathrm{O/H})=8.5$ to $\rho = -0.25$ at $12+\log(\mathrm{O/H})=9.1$. Gradients in the average star formation efficiency in galaxies may lead to deviations in the N/O ratio at fixed O/H. This mechanism does not explain all of the variance in $\xi$ at fixed O/H. Other processes such as the radial migration of stars in spirals or the accretion of passive dwarf galaxies will also influence the N/O-O/H relation, but we are unable to quantify this with our current data. Nevertheless, the empirical effect shown in Figure \ref{OH_NO_RPs} remains. If not taken into account, this may result in biases in the oxygen abundance gradients measured in resolved spectroscopic data.

While the differences in the oxygen abundance gradients shown in Figure \ref{Gradient_differences} was illustrated for an [OIII]/[NII]-based callibrator, we emphasise that this discrepancy can be generalised to other O/H indicators that are contingent on a fixed relationship between N/O and O/H. This includes, but is not limited to, N2S2H$\alpha$ \citep{Dopita2016}, N2O2 \citep{Kewley2002} and O3N2 \citep{Pettini2004}. 

The existence of a universal oxygen abundance gradient over a range of stellar masses has been reported by some authors \citep[e.g.][]{Sanchez2012,Sanchez2014}, though others have found that the gradients vary systematically with total mass \citep[e.g.][]{Belfiore2017, Schaefer2019}. Our work shows that the use of nitrogen-sensitive O/H indicators may have the effect of diminishing the correlation of these gradients with mass due to nitrogen enhancement in the outskirts of massive galaxies. For a more in-depth discussion of the universality of metallicity gradients, see \cite{Sanchez2019}.

Future empirical calibrations of oxygen abundance from a relationship between N/O and O/H will require careful consideration of how HII regions are selected. We caution that the physical conditions, both past and present, will affect this relationship. For detailed studies of oxygen abundance gradients, we recommend the use of an indicator that is independent of N/O or incorporates a correction \citep[e.g.][]{PerezMontero2014,Morisset2016,Pilyugin2016}.

\acknowledgements 
ALS, ZJP and CT acknowledge NSF CAREER Award AST-1554877. 
This research made use of \texttt{Astropy}, a community-developed core \texttt{python} package for astronomy \citep{Astropy2013,Astropy2018} and \texttt{matplotlib} \citep{Matplotlib}, an open-source \texttt{python} plotting library.

Funding for the Sloan Digital Sky Survey IV has been provided by the Alfred P. Sloan Foundation, the U.S. Department of Energy Office of Science, and the Participating Institutions. SDSS acknowledges support and resources from the Center for High-Performance Computing at the University of Utah. The SDSS web site is www.sdss.org. SDSS is managed by the Astrophysical Research Consortium for the Participating Institutions of the SDSS Collaboration including the Brazilian Participation Group, the Carnegie Institution for Science, Carnegie Mellon University, the Chilean Participation Group, the French Participation Group, Harvard-Smithsonian Center for Astrophysics, Instituto de Astrof\'{i}sica de Canarias, The Johns Hopkins University, Kavli Institute for the Physics and Mathematics of the Universe (IPMU) / University of Tokyo, the Korean Participation Group, Lawrence Berkeley National Laboratory, Leibniz Institut f\"{u}r Astrophysik Potsdam (AIP), Max-Planck-Institut f\"{u}r Astronomie (MPIA Heidelberg), Max-Planck-Institut f\"{u}r Astrophysik (MPA Garching), Max-Planck-Institut f\"{u}r Extraterrestrische Physik (MPE), National Astronomical Observatories of China, New Mexico State University, New York University, University of Notre Dame, Observat\'{o}rio Nacional / MCTI, The Ohio State University, Pennsylvania State University, Shanghai Astronomical Observatory, United Kingdom Participation Group, Universidad Nacional Aut\'{o}noma de M\'{e}xico, University of Arizona, University of Colorado Boulder, University of Oxford, University of Portsmouth, University of Utah, University of Virginia, University of Washington, University of Wisconsin, Vanderbilt University, and Yale University.

\bibliography{bibliography}

\label{lastpage}

\end{document}